# Retrieving the Baby: Reichenbach's Principle, Bell Locality, and Selection Bias

Huw Price
Trinity College, Cambridge

**Abstract:** In his late piece 'La nouvelle cuisine' (Bell 1990), John Bell describes the steps from an intuitive, informal principle of locality to a mathematical rule called *Factorizability.* This rule stipulates that when possible past causes are held fixed, the joint probabilities of outcomes of spacelike separated measurements, conditional on measurement settings, be the product of the local conditional probabilities individually. Bell shows that Factorizability conflicts with predictions of QM, predictions since confirmed in many experiments. However, Bell warns his readers that the steps leading to Factorizability should 'be viewed with the utmost suspicion'. He says that 'it is precisely in cleaning up intuitive ideas for mathematics that one is likely to throw the baby out with the bathwater' (2004, 239). Bell's suspicions were well-founded, for he himself misses an important baby. Here we retrieve and identify it: it is *selection bias.* We explain how failure of Factorizability may be regarded as a selection artefact, requiring no violation of locality in the intuitive, conceptual sense with which Bell begins his analysis. The argument begins with a central principle of causal discovery, Reichenbach's Principle of Common Cause (PCC). It is well known that correlations due to selection bias are not subject to PCC. Several writers have proposed that EPR–Bell correlations are also an exception to PCC, but it has not been noticed that they fall under this well-known exclusion. The point is relevant not only to the status of Bell nonlocality, but also for statistics and causal modeling. For these fields, the news is that selection effects play a ubiquitous role in quantum phenomena, in a form akin to collider bias.

## 1. Bell's missing baby

In his late piece 'La nouvelle cuisine' (Bell 1990), John Bell describes the steps from an intuitive, informal principle of causal locality to a mathematical rule called *Factorizability.* This rule stipulates that when possible past causes are held fixed, the joint probabilities of outcomes of spacelike separated measurements, conditional on measurement settings, be the product of the local conditional probabilities individually. Bell shows that Factorizability conflicts with predictions of quantum mechanics (QM), predictions since confirmed in many experiments. So, *modulo* the assumptions of Bell's argument, Factorizability fails. This is the basis for claims of nonlocality in QM, insofar as they stem from Bell's work.

However, Bell warns his readers that the steps leading to Factorizability should 'be viewed with the utmost suspicion'. He says that 'it is precisely in cleaning up intuitive ideas for mathematics that one is likely to throw the baby out with the bathwater' (2004, 239). It turns out that Bell's suspicions were well-founded, for he himself discards an important baby. Here we retrieve and identify it: it is *selection bias.* We explain how failures of Factorizability in EPR–Bell experiments may be represented as selection artefacts, requiring no violation of locality in the intuitive, conceptual sense with which Bell begins his analysis.

Our argument exploits a well-known loophole, or exclusion, to a familiar principle of causal inference. We begin by introducing this principle and the issue of its limits. We then turn to an outline of selection bias in general, before returning to its application in the quantum case.

2. Reichenbach's Principle and its limits

Correlation is not causation, but they keep close company. A central principle of causal discovery tells us that where we find correlation we should expect causation, too – not necessarily *direct* causation between the correlated variables, but otherwise mutual links to other variables, causing both. This is called *Reichenbach's Principle,* or the *Principle of Common Cause* (PCC). David Papineau formulates it like this:[1]

> If A and B are correlated, then one must be causing the other or they must have one or more common causes, and in the latter case controlling for the common causes will screen off the correlation. (Papineau 2025, 9)

This principle is well known in science, though not always under this name. As (Myrvold et al 2024) note, Bell himself relies on a version of it, in formulating his notions of local causality. Writing about Bell's discussion in (Bell 1976), for example, these authors say that implicit in it is the assumption

> that correlations between two variables be susceptible to causal explanation, either via a causal connection between the variables, or via a common cause. This assumption was stated explicitly in a later article (Bell 1981), in which he says that "the scientific attitude is that correlations cry out for explanation" [Bell 1981, 152]. (Myrvold et al 2024, §3.1.1)

Myrvold *et al* identify this assumption with PCC.

Useful though it is, PCC comes with some caveats and exclusions. One exception, according to some writers, concerns the EPR–Bell correlations themselves. David Papineau again:

> EPR correlations certainly pose a challenge to Reichenbach's Principle. After all, they are genuine correlations, by anybody's counting. So, by Reichenbach's Principle, either one measurement is causing the other, or they have a common cause. But in the former case special relativity would require the two measurements not to be spacelike separated, which they aren't, and in the latter case Reichenbach's Principle says that the common cause should screen off the correlation, which Bell's inequality tells us it can't. So, whichever way we turn it, Reichenbach's Principle is in trouble. (2025, 9–10)

Similarly from (Brown & Timpson 2016), describing an option that they take PCC to miss:

---

[1] Papineau distinguishes PCC from Reichenbach's Principle, the former being a corollary of the latter, but the difference won't be significant here.

> [I]n a non-separable theory there is a *further* way in which correlations can be explained which Reichenbach's stipulations miss out: correlations between systems (e.g., the fact that certain correlations between measurement outcomes *will* be found to obtain in the future) can be explained directly by irreducible relational properties holding between the systems, relational properties themselves which can be further explained in dynamical terms as arising under local dynamics from a previous non-separable state for the total system. Which is precisely what happens in the Everettian context, for example. (Brown & Timpson 2016)

Not everyone agrees. Some think that there are direct causal connections in EPR–Bell experiments, despite relativity (e.g., Maudlin 2011, 2014). If so, then Bell correlations fall under PCC after all, even if in a surprising fashion.[2]

I am going to argue that while Bell correlations do count as exceptions to PCC, they don't need their own loophole. They can be handled under the most familiar exception to PCC, so-called *selection bias.* If correct, this makes EPR–Bell phenomena less unusual than they have seemed. It avoids both the challenge of fitting them to PCC and the need for a novel exemption to this central principle of causal discovery.

In other words, I'm going to argue that Bell correlations are *selection artefacts.* I stress that I don't mean that Bell correlations and entanglement are *not real* – that's not in question here. 'Artefact' is simply a familiar label for patterns arising from selection processes. I'm proposing that Bell correlations fit this model. If so, then we learn something about these puzzling phenomena by representing them as selection bias; and we learn something of interest to Reichenbach's contemporary followers, by seeing that selection plays this ubiquitous role in the quantum world.

Is nonlocality real, according to this proposal? That depends on what we mean by the term. We'll return to this in §12, but looking ahead, the upshot will be that if nonlocality means a failure of Bell's condition of *Factorizability,* the proposal confirms and explains it. If it means *direct spacelike influence,* the proposal avoids it – while explaining, *contra* Bell's own discussion, why these two options don't amount to the same thing.

3. Selection bias

What is selection bias? Let's start with a familiar example. Many readers will recognise the image in Figure 1. It represents the distribution of damage found on bombers, returning from missions in WWII. (Think of the red dots as bullet holes.) Clearly, there is a strong correlation between location on the aircraft and the probability of damage at that location. In some parts of the aircraft the probability is high, in other places close to zero.

---

[2] I don't mean that these are the only options. For discussion, see (Myrvold 2016) and (Myrvold et al 2024), from the side of Bell's Theorem, and (Hitchcock & Rédei 2021) from the side of PCC.

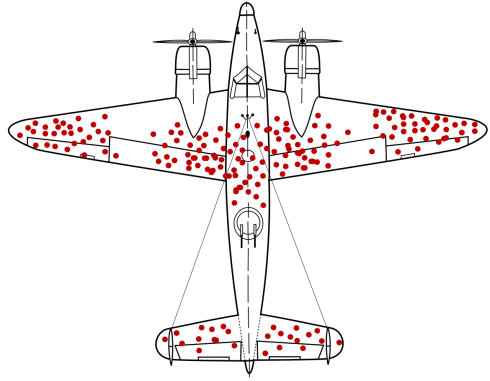

Figure 1: Survivorship bias.[3]

We can imagine ways in which this correlation might be explained via PCC. There are three possibilities.

1. The trajectory of the projectiles causing the damage might be influenced by the geometry of the aircraft – perhaps the enemy gunners reliably aim for those places, or perhaps the aircraft have force fields, steering the bullets to those points and not others.

2. The projectiles might be influencing the location of the aircraft, the latter shifting itself to ensure that the damage is confined to these areas. (Imagine Wonder Woman dodging bullets.)

3. Some third entity might be influencing the location both of the aircraft and of the projectiles, to ensure that the impacts have this distribution. (Imagine an embroiderer, controlling a needle with one hand and the position of the piece of work with the other.)

As we know, none of these explanations is correct. The right explanation, as the Columbia statistician Abraham Wald apparently pointed out,[4] is that the correlation results from the fact that we are looking at a biased subset of the data: namely, those aircraft that made it back to base. Aircraft with damage in other locations did not survive, and that's what's skewing the data, producing the correlation.

This is an example of selection bias. This particular variety is called *survivorship bias,* for the obvious reason: the selection variable is 'survival'. Survivorship bias is a subspecies of so-called *collider bias.* In the terminology of causal models, a collider is a variable influenced by two or more causes – a node where two or more arrows 'collide', in the graphical notation of direct acyclic graphs (DAGS; see Figure 2). Conditioning on a collider – that is, selecting cases in which the collider variable takes a particular value – typically produces a correlation between the contributing causes, even if they are in fact independent. (Again, think of the bomber example.)

---

[3] Image by Granjean and McGeddon from https://en.wikipedia.org/wiki/Survivorship_bias
[4] The story may be apocryphal: https://www.ams.org//publicoutreach/feature-column/fc-2016-06.

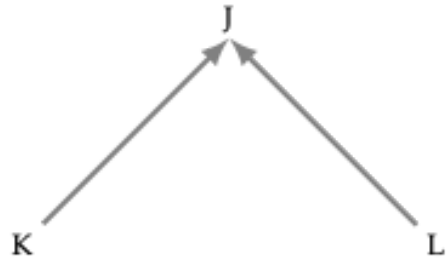

Figure 2: A collider.

Papineau lists 'Collider Bias Correlations', alongside EPR–Bell correlations, as one of three apparent exceptions to PCC.[5] As he says, these correlations 'arise when we control for the common effect of two joint causes in a "collider" structure' (2025, 204). I want to amalgamate these two exclusions into one. In other words, I want to propose that EPR–Bell correlations are similar to collider bias – indeed, they, or some of them, may actually *be* collider bias. But it will be helpful to start further back, stressing first that not all selection bias is collider bias. We'll then be able to state the main proposal as the claim that EPR–Bell correlations arise from selection bias, rather than specifically from collider bias.

4. Selection bias without colliders

Why is selection bias not always collider bias? Three reasons. First, there are non-physical (and hence non-causal) cases of selection bias, e.g., in mathematics. Imagine a study of the correlation between the parity (oddness or evenness) of successive digits of the decimal expansions of two irrational numbers, say $\pi$ and $e$. Let $\pi_n$ and $e_n$ be the $n$th digits in these expansions, respectively. For each $n$, there are four possibilities: the two digits $\pi_n$ and $e_n$ may be even–even, even–odd, odd–even, or odd–odd. Let's assume that these cases are equally likely. This means that there's nothing that the parity of $\pi_n$ tells us about that of $e_n$ and vice versa. (If that's false about $\pi$ and $e$, we can choose different numbers.) Now let S = $\{n \in \mathbb{N} : \pi_n$ is even or $e_n$ is even$\}$ be the set of natural numbers such that $\pi_n$ and $e_n$ are not both odd. Within S, $\pi_n$ and $e_n$ are highly correlated: if one is odd then the other is even. This is a matter of logic, with no causation in the picture. It is purely mathematical selection bias, in other words.[6]

The second reason is subtle, but easy to state, now that we have the mathematical case in view. In some contexts, we might wish to describe a system of real, physical variables in non-causal terms. It is sometimes suggested that there is no true causality in microphysics, for example.[7] It would be foolish to tie our hands on this question, by refusing to contemplate models without causation. In such models, as we'll see, it may still be helpful to talk about selection bias. It won't be collider bias, because *collider* is a causal notion.[8]

---

[5] The third involves the correlation between pairs of time series – not relevant here, so far as I can see. See also discussion in (Hitchcock & Rédei 2021).
[6] We might prefer to withhold the term *selection* here, on the grounds that there is no actual, physical selection involved. The case is useful, nevertheless, in illustrating the simple logical basis that underlies physical cases.

The third reason is not subtle at all, and does apply in the familiar, causal world. A well-known property of common causes is that the correlations they produce disappear if we 'condition on the common cause' – that is, select the cases in which the common cause variable takes one particular value. The explanation is simply that if A co-varies with B because they both co-vary with a common cause C, then if we remove the variation of C, holding it fixed, we remove the co-variation between A and B.

Here's an example. Suppose that rare diseases A and B are both caused in part by possession of a Y chromosome. Perhaps they are both genetic diseases, caused by different abnormalities on the Y chromosome. Few people with a Y chromosome get these diseases, but no one without one does so.

Now suppose that we don't yet know these facts, and we're interested in the question whether these rare diseases are correlated. We investigate the question by studying patient records from our local clinic. If our clinic only treats people with a Y chromosome, then we'll miss the correlation, because we're holding fixed the common cause. Again, this is selection bias, though here it is removing a correlation, not producing one.

This last kind of error is also called *range restriction*.[9] Unlike collider bias, it comes in two forms, depending on when and how the range is restricted. As we described it, we could call it *preselection bias,* because it results from a restriction in advance of the patients admitted to the clinic. But we could imagine the same error as *postselection* bias – suppose now that the clinic treats everyone, but that our algorithm for selecting records only picks those of patients with a Y chromosome.

Summing up, both correlation and lack of correlation can be a result of selection bias. Not all selection bias lives in the world of cause and effect – there are purely mathematical examples, and there may be examples in non-causal models of physical phenomena. When it does live in the physical world, the selection in question may occur either *before* or *after* the creation of the relevant subensemble of cases. The subensemble may be created with the selection built in, in other words, as in the clinic that only admits patients with a Y chromosome. Or the extraction of the subensemble from a larger ensemble may happen 'after the fact', as when the algorithm picks out only those records that list a Y chromosome.

---

[7] Papineau: '[The] affinity with statistical mechanical accounts of thermodynamic behaviour argues that causation is an essentially macroscopic phenomenon, depending on structures which emerge only with respect to coarse-grained aspects of physical systems' (2025, 14).
[8] Objection: selection is a causal notion, at least in real physical cases. Reply: yes, but it is a macroscopic, operational notion – it doesn't guarantee us causation in the micro realm.
[9] See (Dahlke & Wiernik 2019). As these authors use the term, collider bias is also an example of range restriction. I am grateful to Julia Rohrer here.

For some kinds of selection bias, preselection makes little sense. In survivorship bias, for example, the selection is not finalised until fate has finished with the non-survivors. Collider bias in general requires postselection, for much the same reason.[10]

What does selection bias involve? At a minimum, presumably, a distinction between a larger super-ensemble and a smaller subensemble, the latter picked out of the former by some sort of selection process. We get selection bias if there are correlations that *differ* between the super-ensemble and the subensemble. Let's call this the *Minimal Selection Model* (MSM). Later, it will be helpful to make things more stringent, focusing on cases in which the super-ensemble is suitably 'natural', and in which selection is achieved by a real physical process, but this will do for now.

We have seen that the difference in correlations involved in MSM can go in either direction: selection may *remove* as well as *produce* correlations. We have also seen that a super-ensemble need not be an ensemble of *actual* cases. At least where preselection is involved, the selection may be one that prevents some of the super-ensemble from coming into existence in the first place. In such cases, much of the super-ensemble consists of possible but unrealised cases – it is partly a *virtual* super-ensemble, as we might say.[11]

5. Bell correlations by postselection (I): classical toy models

With these preliminaries in place, we are interested in whether Bell correlations can be explained as selection bias. I want to get there in several steps. Starting gently, let's observe that postselection makes it easy to produce Bell correlations in a *classical* toy model.

Suppose that Alice and Bob each choose a 'setting' bit, and toss a coin to generate an 'outcome' bit. They pass the resulting ordered pairs to Charlie, giving him an ordered 4-tuple (a,A,b,B). Charlie discards some of these 4-tuples and retains others, using a probabilistic algorithm to make the cut. The probability that his algorithm will retain a particular result (a,A,b,B) is set to P(A,B|a,b), where the latter is the probability predicted by QM in a standard Bell test of the CHSH inequalities, using an entangled pair of photons in the singlet state.[12]

Trivially, this procedure ensures that in the long run, the correlations in Charlie's ensemble of retained results match those predicted in the real Bell experiment. Equally trivially, it is a case of collider bias. The probability that a result will be retained ('survive') depends both on Alice's two bits and Bob's two bits, so we have a collider. And the correlations emerge when we hold fixed this common effect, ignoring the discarded cases. For future reference, note that Alice and Bob can be spacelike separated. As a selection artefact, the correlation between them poses no challenge to special relativity.

---

[10] Unless causation sometimes works from future to past – we'll come back to that idea.
[11] For a familiar additional example of preselection from a virtual ensemble, see fn. 18.
[12] In familiar notation, a and b are the settings chosen by the experimenters at **A** and **B**, respectively, and A and B are their measurement outcomes.

Let's make this classical toy model a little more sophisticated. The real Bell experiment we had in mind can be performed with any of the four so-called Bell states as the initial preparation. Let's write these four states as $C_0$, $C_1$, $C_2$, $C_3$, where $C_0$ is the singlet state. For each of these states, there is a probability distribution $P(A,B|a,b,C_i)$ giving the predicted joint probability of outcomes (A,B) given settings (a,b) and initial state $C_i$.

Charlie can use these probabilities to sort results from Alice and Bob into four hoppers $H_0$, $H_1$, $H_2$, $H_3$, such that (i) no results are discarded and (ii) the frequencies in each hopper match the predicted results of a Bell test with the corresponding initial state. One way to see that this is possible is to imagine the real Bell experiment performed with the initial state $C_i$ chosen at random, with equal probability for the four options. The inverse probabilities $P(C_i|a,b,A,B)$ are the probabilities needed by Charlie's algorithm, for sorting a result into hopper $H_i$. The results within each hopper exhibit Bell correlations, as before, but again, these are selection artefacts.

6. Bell correlations by preselection (I): a quantum toy model

Leaving the classical toy model behind, consider the QM model just described – a Bell experiment with the initial Bell state chosen at random on each run, with equal probability for each of the four possibilities. This device generates an ensemble of results in which (i) there are no Bell correlations in the ensemble as a whole, but (ii) there are such correlations in each of the four sub-ensembles defined by the initial state $C_i$. This is what guaranteed that the same would be true in the four-hopper version of our classical toy model.

On the face of it, then, we've met the MSM criteria for regarding *some* Bell correlations as selection artefacts, in a real QM case. In a case in which selection produces correlations, MSM calls for two things: an uncorrelated super-ensemble, and a selection procedure that picks out correlated subensembles of this super-ensemble. In the case just described, this selection is performed by the random device that controls the initial state $C_i$. Bell correlations emerge in the subensembles defined by a fixed value of $C_i$.

Thus we have a toy model using real QM components, in which it is easy to see how Bell correlations can be represented as selection artefacts. If we could argue that the toy model is actually *general* – a good representation of *arbitrary* Bell tests with these components – then we would have something interesting. We'd have an argument that such Bell correlations evade PCC *because* they are selection artefacts.

For the moment, though, we just have the toy model. In this model, the uncorrelated super-ensemble is constructed artificially, by an imposed random choice of the initial Bell state. Why not regard this as gerrymandering, rather than a construction with application in the real world? I want to address this concern by going back to the future, to Bell tests with postselection.

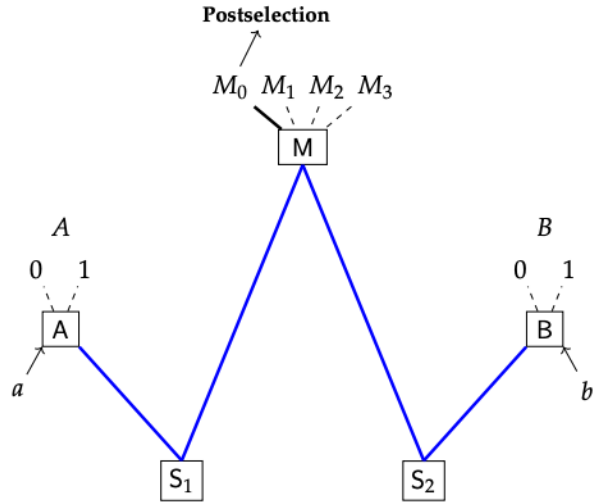

Figure 3: A W-shaped Bell experiment with entanglement swapping

7. Bell correlations by postselection (II): quantum cases

Figure 3 depicts an important experimental protocol, used for some of the best tests of Bell correlations. We can call it the W-shaped protocol, referring to the shape of its arrangement in spacetime. It relies on so-called *entanglement swapping.* Two pairs of entangled particles are produced independently, one pair at the source $S_1$ and the other at the source $S_2$. One particle from each pair is sent to a measurement device at M, which performs a Bell state measurement on the two particles – a measurement with four possible outcomes, shown in Figure 3 as $M_0$, $M_1$, $M_2$ and $M_3$.

The remaining two particles go to measuring devices at **A** and **B**, respectively. The experimenters each choose a setting and record an outcome, in the usual way. The expected results are as follows. There are no **A**–**B** Bell correlations in the ensemble of results as a whole, but such correlations emerge if we postselect on any of the four outcomes at M – e.g., on $M_0$, as shown in Figure 3. As noted, some of the best real Bell experiments use this protocol.

It is easy to see that the case meets our MSM criteria for the presence of selection artefacts. The super-ensemble is the set of results as a whole, and the sub-ensembles are the four sets of results we obtain by postselection at M. In the sense implied by MSM, Bell correlations in such experiments simply *are* selection artefacts – there's no question about the matter.[13]

Readers may protest that there is real entanglement between the two sides of such an experiment – it is certainly not an artefact. This is an opportunity to re-emphasise our terminology. As I said at the end of §2, the term 'artefact' is not intended to challenge the *reality* of entanglement in such cases. It is intended to offer a *diagnosis* of it, with respect to the options allowed by PCC and its known exceptions. The diagnosis is that the correlations in question are produced by selection. This means both that they don't need to be explained within the scope of PCC, by direct causal links or common causes; and that they don't need any new kind of exception to PCC – they fall under a well-known exception.

---

[13] There is a question whether they are *just* selection artefacts – see §10.

Two more objections. First, even if the diagnosis is accepted for the Bell correlations 'across the W' from **A** to **B**, it says nothing, at this point, about the entanglement in the two wings of the experiment, needed to produce the results at M that in turn yield postselected correlations between **A** and **B**. This objection is fair, and for the moment let's set it aside, with the following note. If we had a diagnosis that worked in ordinary two-particle ('V-shaped') Bell experiments, then we could apply it here, to deal with the wings as well. The W experiments would then be regarded as hybrid cases, involving selection at all three vertices.[14]

Second, our diagnosis has said nothing about an important feature of the geometry of these W-shaped experiments, a feature that makes a big difference to how they are represented in orthodox QM. The measurement M can have three possible locations with respect to those at **A** and **B**.[15] It may be in their absolute future – so-called *delayed choice* entanglement swapping (DCES) – in their absolute past, or somewhere in between. In the orthodox QM picture with collapse, this makes a big difference to the description of the case. In the DCES case the measurements at **A** and **B** collapse the state of the second particle in each pair, influencing the states reaching M. In the absolute past case, the influence goes the other way: the measurement at M affects the states of the particles received at **A** and **B**.

In the DCES case, as several authors have noted, this means that we have something like a causal story, for how information at **A** and **B** can make a difference to the outcome at M. It becomes plausible to regard M as a collider, in other words, and hence, perhaps, to treat the resulting **A**–**B** correlations as selection artefacts – a case of collider bias, in fact.[16] But this story won't work, apparently, in the other possible W geometries.

This is why I stressed earlier that not all selection bias is collider bias. If we were trying to force Bell correlations into a box labeled *Collider Bias* then the geometry of the W cases would certainly be a concern. (More on this below.) But MSM doesn't mention causal structure, merely statistical structure: namely, a difference in the correlations within the super-ensemble and selected sub-ensembles. By this purely statistical criterion, it seems indisputable that the Bell correlations are selection artefacts, in all the W geometries.[17]

---

[14] Why wait? We can set up the case with four possible inputs at $S_1$ and $S_2$, which gives us a super-ensemble with $4^3$ possibilities. After that, I think the argument runs as before.

[15] I'm ignoring some hybrid possibilities, e.g., cases in which **A** and **B** are not spacelike separated from one another.

[16] See (Gaasbeek 2010), (Egg 2013), (Fankhauser 2019), (Bacciagaluppi & Hermens 2021), (Price & Wharton 2021), and (Mjelva 2024).

[17] Again, it is not indisputable that they are *just* selection artefacts; see §10.

As I've already emphasised, the term *artefact* is not intended as a challenge to the reality of the entanglement correlations in these cases. The diagnosis simply removes the pressure to construe these correlations in causal terms, insofar as it stems from PCC.

8. Bell correlations by preselection (II): removing the gerrymander

In §6 we described a toy model involving a two-particle ('V-shaped') Bell experiment. We imagined the initial Bell state chosen at random, by some suitable device. Our reason for not regarding this toy model as generalisable to real V-shaped Bell experiments was that it seemed to involve a gerrymander in the construction of the super-ensemble: in effect, we defined the super-ensemble by simply gluing together four distinct possibilities, with equal weight.

We have seen that in W-shaped versions of a similar experiment – it can be so similar that Alice and Bob, confined in their laboratories, do not know in which version of the experiment they are participating – no gerrymandering is necessary. The QM rules naturally produce the four possible outcomes at M, with equal probability. At this point I'll invoke what I'll call the *Time-Symmetry Conjecture* (TSC), to propose that we should regard these same probabilities as the natural measure for the initial Bell state in the V-shaped version of the experiment – in other words, the measure that would apply if the initial state were not controlled by an experimenter or some other control device.

It would take a long detour to defend TSC in detail, though the path that detour would take is well-marked. In the standard modern version of the Boltzmannian orthodoxy, all familiar macroscopic time-asymmetry results from the imposition of a low entropy boundary condition in the distant past (the so-called Past Hypothesis). The detour would take us to an imaginary regime, easily found in this orthodoxy, in which there is no such boundary condition. In this regime, there's no basis for a difference between forwards and backwards predictions in QM. The probabilities of 'inputs' to the V-shaped experiment would therefore match those of 'outputs' at M in the W case. For present purposes, as I say, it would be a distraction to make this case in detail. I'll simply assume it, in the form of TSC.[18]

TSC removes the gerrymander. In other words, it explains why what we postulated explicitly in the V-shaped toy model in §6 – the four inputs with equal probability – is actually the natural case, found on Nature's shelf. So we don't need to put it in by hand. Instead, we can simply observe that the familiar case, where we pick an initial Bell state of our choice, amounts to a preselection from a natural (virtual) super-ensemble. We have the ingredients for MSM, in other words, with no need for anything artificial.

---

[18] For details, see (Price & Wharton 2025). Note that by the lights of MSM, the familiar time-asymmetry of thermodynamics, and everything that depends on it, is itself a preselection artefact. The role of the Past Hypothesis is to select a tiny, time-asymmetric subensemble from a vastly larger, virtual, time-symmetric super-ensemble.

## 9. Where we are

We've shown that Bell correlations have a natural representation as selection artefacts. This means that there is no longer pressure from PCC to fit them under the two headings that Reichenbach allows: direct causation or co-variation due to common causes. In that sense, the proposal gets us off a large hook, without additional cost – without needing to summon up a *new* exclusion to PCC, in other words.

Moreover, as promised at the beginning, there's significant news here for Reichenbach's descendants. It is well known that V-shaped forks (i.e., systems of two variables jointly linked to a third) come in two varieties. There are (i) what Reichenbach called *conjunctive forks*, in which conditioning on the vertex variable *de-correlates* the other two variables; and (ii) what we could call *correlating forks*, where conditioning does the opposite. In typical examples, the former involves a common *cause* in the common *past* of the other two variables, the latter a common *effect* (collider) in their common *future*.

The discussion above shows that this conventional picture needs a large addendum. EPR-Bell cases show that correlating forks with the opposite temporal orientation are extremely common. Again, this doesn't seem to be a contestable view, at least across a wide range of views of quantum ontology.[19] From most points of view, the facts on which the proposal depends are simply *there* in the operational correlations of QM. It is often observed that entangled preparation states do not behave like common causes – fixing them does not screen off the correlations downstream. However, it doesn't seem to have been noted explicitly that there's a sense in which they do the opposite. With respect to their behaviour under conditionalisation, they work like common effects, or colliders.

Are these correlating forks in QM best regarded as *actual* colliders, or do they merely behave like colliders in the respect we have described? This turns out to be a subtle matter. That's why it was helpful to step back at the beginning, discussing *selection bias* rather than *collider bias* specifically. If we hadn't stepped back in this way, we would have left ourselves open to the charge that we hadn't proved that there are really colliders involved. As we actually proceeded, our conclusions don't depend on this question – they are not hostage to a causal model. But the question is interesting, nevertheless, especially for causal modellers. What can we say?

It turns out that there's another fact about selection bias and PCC that it will be helpful to have in view, before we return to this question about colliders. I'll explain it using some simple examples, classical and quantum.

---

[19] The apparent exceptions are views that are not sufficiently realist about the operational facts – see, e.g., (Smerlak & Rovelli 2007) and (Müller 2020) – and Everettian views, in which these facts are embedded in a much larger ensemble.

## 10. Selection bias doesn't *prohibit* PCC-compatible causation

Representation of a correlation as a selection artefact does not *prove* that PCC is inapplicable. It frees us from an *obligation* to meet the demands of PCC, but it is not a decisive obstacle to someone who proposes to meet those demands in any case. I'll give three examples.

First, think of the bombers. We can imagine a world in which bombers are equipped with force fields, steering incoming projectiles to particular parts of the aircraft. Perhaps these force fields have two settings – one steers the bullets to the regions shown in Figure 1, the other to the complementary regions. As it turns out, the latter setting is disastrous – those planes do not survive. Then the distribution shown in Figure 1 is still survivorship bias, but in the surviving aircraft it also has a PCC explanation: the planes influence the trajectory of the bullets.

Second, here's a simple classical example with common causes. Imagine a factory minting coins. The production line has two settings. One produces ordinary coins with a Head on one side and a Tail on the other. The other produces coins with the same thing on both sides – either two Heads or two Tails, with equal probability. The choice between the settings is made at random, at the beginning of each shift.

Alice and Bob are quality controllers. They look at opposite sides of each coin, as it emerges from the production line. Clearly, what they see is either correlated or anticorrelated, depending on the factory setting at the time. But they can't tell which it is, if they only have access to their own records.

If we look at the results overall, there is no correlation between Alice's and Bob's records of the same coins. But there is a correlation in the subensembles corresponding to each of the factory settings. (Perhaps it isn't 100%, because Alice and Bob sometimes make mistakes, but that doesn't matter here.) How should we explain the correlation in these subensembles? Unless the precise wording of the question pushes us one way or the other, we simply have a choice. The setting preselects a correlated subensemble; and, whichever subensemble it is, there's a common cause in the factory process, responsible for the correlation in question.

Third, the same seems to be true in QM. Think of a Bohmian, for example, who offers us a model in which there is spacelike causality, or something close to it. The representation of EPR–Bell correlations as selection artefacts, in the sense of MSM, seems no less applicable in the Bohmian model than in others. The representation is entirely operational, after all. It may remove some *motivation* for a Bohm model, but it isn't *incompatible* with it.

The general lesson is this. Selection bias can remove the *need* for a causal explanation of the kinds described by PCC, but it doesn't prove the *absence* of such causation. Applying this lesson to the quantum case, it means that the representation of Bell correlations as selection artefacts is not in itself an argument that PCC is inapplicable. In this case, however, we have strong independent reasons for doubting whether the case fits PCC. This is where we came in:

Bell's Theorem and the experimental violation of Bell inequalities count against common causes, while relativity counts against direct causation. The significance of the diagnosis of Bell correlations as selection bias is that it gives us an additional option, and one that doesn't require any *new* exclusion to PCC. Selection bias is a well-known exception.

11. Searching for colliders

Now to our question about colliders. Bell correlations can be represented as selection bias, but is it necessarily *collider* bias? One way to address this question is to try to 'work from within', looking for mechanisms to explain the operational correlations on which the application of MSM depends. The search for such mechanisms seems a worthy goal, but can it give us the colliders we would need, to map the proposal from selection bias to collider bias?

The prospects seem bleak. There are a number of different mechanisms possible, not all of which give us colliders where we would need them. We can see this in the W cases, where, as we noted, different locations of M with respect to the measurements at **A** and **B** are treated differently in the orthodox model. M may well be a collider in orthodox QM in DCES, but the story is different when M is in the common past of **A** and **B**. So the search for a general mapping is already in trouble, before we get to the two-particle V-shaped cases. We may have objections to the orthodox model, but it is a counterexample to any proposal to *necessitate* colliders, by adding ontology to our operational model.

Instead, I propose that we work from the outside. We look at the operational structure, starting with the geometries in which colliders seem most plausible, and then use the operational symmetries to argue that in this thin operational sense, we have them in all cases – whatever the ontology.

Focusing on the DCES case, the orthodox model makes (a,A) and (b,B) available at M, because the collapsed states of the two incoming left and right particles reflect this information, respectively. So (a,A) and (b,B) both influence the *probabilities* at M, just as in our classical toy model. It also seems natural to talk of these pairs as *making a difference* at M, in a counterfactual sense – if one or both had been different, the outcome at M might well have been different. This argument could be strengthened by using a case that allows for the same settings in the measurement devices at **A** and **B**. That gives us access to the perfect correlations or anti-correlations of EPR cases, meaning that some joint values of (a,A) and (b,B) are inconsistent with some outcomes at M. That, in turn, means that we can find non-probabilistic counterfactuals, to the effect that certain differences on one side or other would *require* a difference at M – which looks like a collider.

Two notes. On the one hand, we got this result by treating both the settings and outcomes at **A** and **B** as contributing factors. We won't get such a strong result if we restrict ourselves to the settings, because in that case the outcomes (A,B) will be an alternative 'sink' where counterfactual changes can make a difference. On the other hand, the orthodox model plays no real role in the argument of the previous paragraph. We have the probabilities and correlations we need from purely operational considerations.

Accentuating the positive, let's live with the first note and run with the second. The operational correlations are the same as they are in the DCES case, in all versions of the W experiments and in the corresponding version of the two-particle V-shaped experiment. Can we use this symmetry to map the argument to those cases?

Short answer: no, because the counterfactuals work differently. By ordinary standards, the past is held fixed in counterfactual reasoning. So changes to (a,A) and (b,B) cannot show up at M in the non-DCES W cases, or at the initial state at C in the V cases. We don't have the counterfactuals needed for a collider in these cases. Shifting to a version of the experiment with EPR correlations won't help matters. It will simply show us that treating the outcomes (A,B) as well as the settings (a,b) as inputs can lead to inconsistencies, where the operational variables are over-constrained.

Long answer: let's think about why the counterfactuals work differently. Again, it would take us a long way off track to discuss this in detail, but there is a considerable degree of consensus that this, too, depends on the Past Hypothesis (PH).[20] If so, then the move proposed in §8 – considering a regime without PH – might be expected to remove the ordinary time-asymmetry in the counterfactuals. The upshot will be that in this regime, we do have colliders in all the W and V cases, in the thin, operational sense described above for the DCES case. In this regime, there is no time-asymmetry to mandate that causation only works from past to future; nor, apparently, to mandate that (A,B) cannot be treated as input variables.

Bell himself was cautious about the use of the term causation in physics. In an early piece he expresses one of his reservations, and his view of its resolution, like this.

> In this matter of causality it is a great inconvenience that the real world is given to us once only. We cannot know what would have happened if something had been different. We cannot repeat an experiment changing just one variable; the hands of the clock will have moved, and the moons of Jupiter. Physical theories are more amenable in this respect. We can *calculate* the consequences of changing <u>free</u> elements in a theory, <u>be they only initial conditions</u>, and so can explore the causal structure of the theory. (Bell 1977, 101; underlining added)

Bell puts a time-asymmetry in by hand here, with the underlined phrase. To discuss causality in a regime without PH, we can simply take it out. We can calculate the consequences of changing other elements, and be liberal about what we count as a free or exogenous element. With these small tweaks to Bell's own conception of causality in physics, we get a time-symmetric picture.[21] In this picture, we get colliders where we want them, in all the experimental geometries.

---

[20] See (Price & Wharton 2025) for discussion and references.

Does this result matter? Not to the main proposal of this piece, which is that Bell correlations can be represented as selection artefacts. As I've stressed, that doesn't depend on whether the selection variables in question can be regarded as colliders. However, if we are interested in exploiting and explaining this result – in taking advantage of the freedom it gives us to ignore PCC, and in looking for an ontological model to understand how the quantum world produces the operational correlations on which these selection artefacts depend – there is much to be said for starting with this bare, time-symmetric picture. The regime without PH is a long way from the familiar world of everyday life, or everyday physics. But if we want to regard QM as a fundamental theory, we want a model that would work in such a regime, and not be hostage to the presence or absence of PH.[22]

This starting point would *allow* us to speak of colliders, in the thin sense just described. Would it *require* that we do so? No, because there will also be the option of avoiding causal terminology altogether, in the initial stages of the project. That would involve the same methodology as our tweaked version of Bell, but without *causality* as a label for the structure we were investigating. An advantage of this cautious approach would be to discourage Reichenbach's well-meaning descendants from looking over our shoulder, eager to share the insights of their causal modeling. Like Bell's own assumption of the time-asymmetry of causation, these 'insights' would need to be examined with care, to make sure, reversing Bell's metaphor, that we were not *importing* a baby with the bathwater.[23] So there's no harm in establishing a bit of distance at the beginning, avoiding the term *causality* while the project is still *in utero.*

12. The meaning of nonlocality

I have argued that Bell correlations may be regarded as selection artefacts. Is this merely a *diagnosis* of nonlocality, or is it a way of avoiding nonlocality altogether? As foreshadowed at the end of §2, it depends on what we mean by the term nonlocality. I'll explain this point with reference to an account of Bell's own discussion by (Myrvold et al 2024). In the following passages, these authors describe Bell's analysis of locality in his late piece 'La nouvelle cuisine' (Bell 1990).

---

[21] The Future-Input-Dependent (FID) models discussed in (Wharton & Argaman 2020) are a step in this direction, though these authors do not go as far as relaxing PH, or treating outcomes as free variables.
[22] Advocates of retrocausal/FID models have often pointed out that it is a defect of discussion of causality in QM by Bell and others that they simply help themselves to the ordinary past-to-future conception of causality, blind to the reasons for thinking that it is not fundamental. See (Wharton & Argaman 2020; Friederich & Evans 2023) for discussion and references.
[23] We'll come to Bell's use of the metaphor in a moment. I have reversed it to remind readers of the risks of taking familiar time-asymmetries for granted, in discussing fundamental matters.

On the basis of … Lorentz invariance and the assumption that causes temporally precede their effects, Bell introduces what he calls the *principle of local causality*.

(PLC-1)   The direct causes (and effects) of events are near by, and even the indirect causes (and effects) are no further away than permitted by the velocity of light.

This, says Bell, is "not yet sufficiently sharp and clean for mathematics." [1990, 239] For this reason, he introduces what he presents as a sharpened version of the principle (refer to Figure [4]).

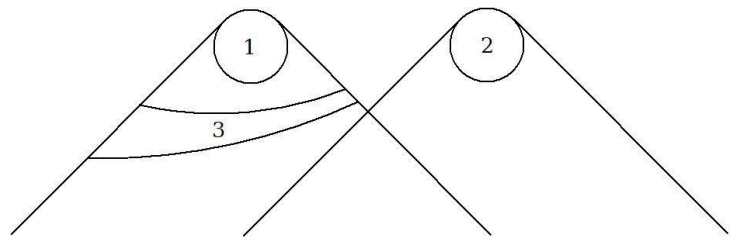

Figure 4 (from Myvold et al 2024)

(PLC-2)   A theory will be said to be locally causal if the probabilities attached to values of local beables in a space-time region 1 are unaltered by specification of values of local beables in a space-like separated region 2, when what happens in the backward light cone of 1 is already sufficiently specified, for example by a full specification of local beables in a spacetime region 3.

The transition from what we have called PLC-1 to PLC-2 should, according to Bell, "be viewed with the utmost suspicion" as "it is precisely in cleaning up intuitive ideas for mathematics that one is likely to throw the baby out with the bathwater" (2004, 239). The relation between them is not further discussed in that article, but remarks in other papers shed light on the transition between them. …

PLC-1 does not, by itself, entail that correlations be causally explicable, and, indeed, does not commit to any sort of causal relations in the world; it merely says that whatever causal relations there are respect relativistic locality. For this reason, we will refer to it as a *causal locality* principle. This is in contradistinction to PLC-2, which requires there to be causal relations of some sort wherever correlations are found. What Bell calls the Principle of Local Causality, PLC-2, can be thought of as a conjunction of (1) a causal locality condition along the lines of PLC-1, restricting causes of an event to that event's past light cone, and (2) Reichenbach's Common Cause Principle [PCC], which requires that correlations be causally explicable. (Myrvold et all 2024, §3.1.1)

This is a clue to where the present proposal diverges from Bell's framework. As we have stressed, selection artefacts are an exception to PCC. They are correlations that do not require explanation via common causes or direct causation. More on this in a moment, but first to the formal condition called *Factorizability,* which Bell takes to follow from the non-formal, conceptual considerations just described. Modifying Bell's notation slightly, we can write this condition as follows:

(FACT)   $P(A,B|a,b,C) = P(A|a,C)P(B|b,C)$[24]

About this condition, Bell says the following.

> Now this formulation has a very simple interpretation. It exhibits A and B as having no dependence on one another, nor on the settings of the remote polarizers (b and a respectively), but only on the local polarizers (a and b respectively) and on the past causes, [C] … We can clearly refer to correlations which permit such factorization as 'locally explicable'. Very often such factorizability is taken as the starting point of the analysis. Here we have preferred to see it not as the *formulation* of 'local causality', but as a consequence thereof. (Bell 1990, 243)

We now have three things that might be meant by Locality. In reverse order, they are FACT, PLC-2 and PLC-1. In which of these senses, if any, is the present analysis a confirmation of *nonlocality* – that is, of a *failure* of Locality? I propose the following answers.

1. The proposal does not conflict with PLC-1. It introduces no spacelike *causation,* direct or indirect.
2. The proposal does conflict with FACT. Selection artefacts can produce non-factorizable correlations between independent variables, and are doing so, in this case. Indeed, FACT amounts to the stipulation that C does *not* behave as what we termed a *correlating fork* in relation to (a,A) and (b,B); and hence is falsified by the confirmation that it does do so.
3. This leaves PLC-2. Here the most plausible reading is that this fails, albeit for a reason that Bell doesn't spot: namely, the relevance of an otherwise well-known exception to PCC, the case of selection artefacts. Bell was thus right to regard the transition from PLC-1 to PLC-2 with suspicion. What he threw out with his bathwater was a well-known and widely-studied statistical baby, most familiar under names such as *collider bias* and *survivorship bias.*[25]

---

[24] The main modification I have made to Bell's formulation is to remove possible 'hidden variables' λ from the conditions of all these conditional probabilities. Our present framework is purely operational, and this simplification makes no difference to the conclusions drawn below.

[25] We might defend PLC-2, and perhaps FACT too, by noting that Bell's discussion concerns beables and their 'ontological' probabilities, and that selection artefacts do not reflect those. But this will merely move the loophole from one place to another. If we include FACT in this response, the upshot will be that because the operational frequencies do involve selection bias, they don't establish that FACT fails in the ontological sense.

## 13. Conclusion

Against the background of Bell's Theorem and experimental confirmation of violation of Bell inequalities, the present diagnosis *saves* Locality in the sense of PLC-1, but *confirms* Nonlocality in the sense of PLC-2 and FACT. In the latter cases, however, the fault lies with the criteria in question (PLC-2 and FACT). Our analysis shows that when the possibility of selection artefacts is acknowledged, there is an obvious loophole in each of these conditions. In each case, the loophole allows for a failure of the criterion in question, without a violation of Locality in the intuitive sense of PLC-1. This is the baby that Bell overlooks.

A final note. Attempts to rescue Locality, in the light of Bell's work and observed violation of Bell inequalities, often focus on an assumption of Bell's Theorem called *measurement independence* or *statistical independence* (SI). This assumption stipulates that the measurement settings (a,b) are independent of hidden variables λ in the overlap of the past light cones of the two measurements. There are two common strategies for challenging SI, sometimes confused. One, *superdeterminism,* proposes that a, b, and λ may all be influenced by common causes in their mutual pasts, and correlated for that reason. The other, *retrocausality,* continues to treat the settings a and b as free or exogenous variables, but proposes that these variables may have effects in their common past (causation being permitted to work backwards, according to these proposals, at least in the micro realm).

The present proposal claims to rescue Locality, at least in the sense of PLC-1. Does it therefore require a challenge to SI? If so, which of these two routes does it take? Answer: No, and neither. We are not challenging Bell's proof that FACT implies Bell inequalities, or the experimental demonstrations that QM violates Bell inequalities. Unchallenged, these together imply that FACT fails, and the present proposal agrees. However, FACT's failure – nonFACTuality, as we might call it – can be understood as a selection artefact, not requiring a violation of Locality in the sense of PLC-1.[26]

## References

Bacciagaluppi, G. & Hermens, R., 2021. Bell inequality violation and relativity of pre- and postselection. arXiv:2002.03935

Bell, J S, 1976, The theory of local beables. *Epistemological Letters*, 9: 11–24. Reprinted in (Bell 2004), 52–62. Page references are to the latter version.

Bell, J S, 1977. Free variables and local causality. *Epistemological Letters* 15, 79–84. Reprinted in (Bell 2004), 100–104. Page references are to the latter version.

---

[26] This piece is greatly indebted to Ken Wharton, and builds on our joint piece (Price & Wharton 2025). I am also grateful to David Papineau, Simon Friedrich and Travis Norsen. I first heard the suggestion that preparation is preselection from Gerard Milburn. It took me a long time to understand what he meant, and discussions with Heinrich Päs and his group, in Dortmund in 2023, played a crucial role.

Bell, J S, 1981, Bertlmann's socks and the nature of reality. *Journal de Physique Colloque C2, supplément au nº 3, Tome 42*: C2-41–46. Reprinted in (Bell 2004), 139–158. Page references are to the latter version.

Bell, J S, 1990. La nouvelle cuisine. In *Between Science and Technology*, A. Sarlemijn and P. Kroes (eds.), 97–115. Amsterdam: Elsevier. Reprinted in (Bell 2004), 232–248. Page references are to the latter version.

Bell, J S, 2004. *Speakable and unspeakable in quantum mechanics*, 2nd edn. Cambridge: Cambridge University Press.

Bell, M. and S. Gao (eds.), 2016. *Quantum Nonlocality and Reality: 50 years of Bell's theorem*, Cambridge: Cambridge University Press.

Brown, H. R., and Christopher G. Timpson, C.G., 2016. Bell on Bell's theorem: the changing face of nonlocality. In Bell and Gao (eds.) 2016: 91–123.

Dahlke, J. A. and Wiernik, B. M., 2019. Not Restricted to Selection Research: Accounting for Indirect Range Restriction in Organizational Research. *Organizational Research Methods*, *23*(4), 717-749. https://doi.org/10.1177/1094428119859398

Egg, M., 2013. Delayed-choice experiments and the metaphysics of entanglement. *Foundations of Physics,* 43, 1124–1135.

Fankhauser, J., 2019. Taming the delayed choice quantum eraser. *Quanta* 8, 44–56. arXiv:1707.07884

Friederich, Simon and Peter W. Evans, 2023.  Retrocausality in Quantum Mechanics. *The Stanford Encyclopedia of Philosophy* (Winter 2023 Edition), Edward N. Zalta & Uri Nodelman (eds.), URL = <https://plato.stanford.edu/archives/win2023/entries/qm-retrocausality/>.

Gaasbeek, B., 2010. Demystifying the delayed choice experiments. arXiv:1007.3977

Hitchcock, Christopher and Miklós Rédei, 2021.  Reichenbach's Common Cause Principle. *The Stanford Encyclopedia of Philosophy* (Summer 2021 Edition), Edward N. Zalta (ed.), URL = <https://plato.stanford.edu/archives/sum2021/entries/physics-Rpcc/>.

Maudlin, T., 2011. *Quantum Non-Locality and Relativity*, Oxford: Blackwell. 3rd ed.

Maudlin, T., 2014. What Bell did. *Journal of Physics A: Mathematical and Theoretical*, 47: 424010.

Mjelva, Jørn, 2024. Delayed-choice entanglement swapping experiments: no evidence for timelike entanglement. *Studies in History and Philosophy of Science* 105, 138–148.

Myrvold, W.C., 2016. Lessons of Bell's theorem: nonlocality, yes; action at a distance, not necessarily. In Bell and Gao (eds.) 2016, 237–260.

Myrvold, Wayne; Genovese, Marco; Shimony, Abner, 2024. Bell's Theorem. *The Stanford Encyclopedia of Philosophy (Summer 2024 Edition)*, Edward N. Zalta & Uri Nodelman (eds.), URL = <https://plato.stanford.edu/archives/sum2024/entries/bell-theorem/>.

Müller, Markus, 2020. Law without law: from observer states to physics via algorithmic information theory. *Quantum* 4, 301.

Price, Huw & Wharton, Ken, 2021. Entanglement swapping and action at a distance. *Foundations of Physics* 51, 105. doi.org/10.1007/s10701-021-00511-3

Price, Huw & Wharton, Ken, 2025. Taming entanglement. https://arxiv.org/abs/2507.15128.

Smerlak, M. & Rovelli, C., 2007. Relational EPR. *Foundations of Physics* 37: 427–445.

Wharton, K. & Argaman, N., 2020. Bell's Theorem and locally-mediated reformulations of quantum mechanics. *Reviews of Modern Physics* 92, 21002. arXiv:1906.04313